# Resolving the Nucleation Stage in Atomic Layer Deposition of Hafnium Oxide on Graphene

Bernhard C. Bayer,[1,2,*] Adrianus I. Aria,[3] Dominik Eder,[1] Stephan Hofmann,[4] Jannik C. Meyer[2,#]

[1]*Institute of Materials Chemistry, Vienna University of Technology (TU Wien), Getreidemarkt 9, A-1060 Vienna, Austria*

[2]*Faculty of Physics, University of Vienna, Boltzmanngasse 5, A-1090 Vienna, Austria*

[3]*Surface Engineering and Precision Institute, School of Aerospace, Transport and Manufacturing, Cranfield University, College Road, Cranfield, MK43 0AL, UK*

[4]*Department of Engineering, University of Cambridge, 9 JJ Thomson Avenue, Cambridge, CB3 0FA, UK*

[*]Corresponding Author e-mail: bernhard.bayer-skoff@tuwien.ac.at

[#]Present address: Institute for Applied Physics, Eberhard Karls University of Tuebingen, Auf der Morgenstelle 10, D-72076 Tuebingen, Germany and Natural and Medical Sciences Institute at the University of Tuebingen, Markwiesenstr. 55, D-72770, Reutlingen, Germany

ABSTRACT

The integration of two-dimensional (2D) materials with functional non-2D materials such as metal oxides is of key technological importance for many applications, but underlying mechanisms for such non-2D/2D interfacing remain largely elusive at the atomic scale. To address this, we here investigate the nucleation stage in atomic layer deposition (ALD) of the important metal oxide $HfO_2$ on chemical vapor deposited graphene using atomically resolved and element specific scanning transmission electron microscopy (STEM). To avoid any deleterious influence of polymer residues from pre-ALD graphene transfers we employ a substrate-assisted ALD process directly on the as grown graphene still remaining on its Cu growth catalyst support. Using this approach we resolve at the atomic scale key factors governing the integration of non-2D metal oxides with 2D materials by ALD: Particular to our substrate-assisted ALD process we find a graphene-layer-dependent catalytic participation of the supporting Cu catalyst in the ALD process. We further confirm at high resolution the role of surface irregularities such as steps between graphene layers on oxide nucleation. Employing the energy transfer from the scanning electron beam to *in situ* crystallize the initially amorphous ALD $HfO_2$ on graphene, we observe $HfO_2$ crystallization to non-equilibrium $HfO_2$ polymorphs (cubic/tetragonal). Finally our data indicates a critical role of the graphene's atmospheric adventitious carbon contamination on the ALD process whereby this contamination acts as an unintentional seeding layer for metal oxide ALD nucleation on graphene under our conditions. As atmospheric adventitious carbon contamination is hard to avoid in any scalable 2D materials processing, this is a critical factor in ALD recipe development for 2D materials coating. Combined our work highlights several key mechanisms underlying scalable ALD oxide growth on 2D materials.

**Introduction**

To realize the envisioned potential of two-dimensional (2D) materials in device applications, the integration of 2D materials with a variety of functional non-2D materials is critically required. Particularly functional metal oxides are technologically highly important non-2D materials to be integrated in 2D/non-2D configurations as e.g. dielectric layers in 2D electronics,[1–3] barrier layers in spintronics[4], environmental encapsulation layers,[5] charge transfer dopants in transparent conductor applications[6] or as photo-catalysts for energy applications.[7]

Metal oxides can be integrated with 2D materials via a variety of processes including evaporation, sputter deposition or atomic layer deposition (ALD) of the oxide onto the 2D material. Key requirements are control over thickness, coverage and microstructure of the oxide, compatibility with other established device processing flows and preservation of the structure of the 2D material onto which the oxide is deposited.[8–10] ALD is most promising towards all these requirements and has therefore emerged as a popular route for metal-oxide/2D integration.[8,9] A remaining key bottleneck of ALD is however often insufficient control over oxide film nucleation on the comparatively inert basal planes of 2D materials. This intrinsically relates to the growth mechanisms in ALD in which (metal-)precursor molecules and oxidant are sequentially pulsed into the chamber. In the initial stages of growth the availability of binding sites on the substrate for precursor and oxidant critically governs ALD oxide nucleation behaviour.[11] Such binding sites are comparatively absent on the 2D materials' basal planes. This often leads for 2D materials to preferential, inhomogeneous nucleation of ALD oxides only at chemically reactive surface irregularities such as steps, folds or defects.[12–15] In turn this can impede formation of pin-hole-free homogeneous oxide films (particularly for low oxide film thicknesses), contrary to what is required for 2D materials applications. To address this, many empirically calibrated processing recipes including seed layers[16–23] and pre-treatment schemes[20,24–29] have been developed. Little work has to date however actually focused on the mechanistic understanding behind nucleation and growth of such ALD metal oxide films on 2D materials. In particular, very little work[21,30–32] has focused on imaging a growing ALD metal oxide film and its 2D material support at high resolution during ALD oxide nucleation. This is in spite of the potentially highly valuable insights towards rational process control that such visualization of the fundamental nucleation and growth steps might provide.

We therefore present here an atomically resolved and element specific scanning transmission electron microscopy (STEM) study of ALD of the important metal oxide hafnium oxide ($HfO_2$) on chemical vapor deposited (CVD) graphene grown on Cu catalyst. $HfO_2$ is frequently used as a high dielectric constant (high-*k*) dielectric[33] in 2D electronics and thus serves as an archetypical metal-oxide model in our study while graphene acts as an archetypical 2D support. For ALD we employ a substrate-assisted ALD[28,34] process that avoids any detrimental influence of residues from polymer-assisted graphene transfers[35] pre-ALD by performing the ALD directly on the as grown graphene films still remaining on their Cu catalyst supports i.e. without any pre-ALD transfer steps involved. We thereby avoid polymer-assisted pre-ALD graphene transfers leading to polymer-residue-governed ALD nucleation.[12,27,36] This allows us to probe an as clean as possible, intrinsic ALD oxide/graphene interface for scalable ALD conditions. Substrate-assisted ALD has been previously shown for various combinations of 2D materials on their growth catalysts and their device integration.[4,5,28,29,34,37,38] The mechanisms behind substrate-assisted ALD have been hypothesized to result from the 2D materials being thin enough for the electronic properties of the catalyst substrate to emanate through the 2D layer during the ALD coating.[34]

This in turn was suggested to lead to the underlying metal catalyst partaking catalytically[39] in the ALD reactions, thereby improving the homogeneity of ALD film growth via this "substrate-assistance" (Fig. 1).

After releasing such grown ALD oxide/graphene stacks from the Cu catalyst,[40] we employ aberration corrected STEM to provide atomic resolution and element specific[41] clarification of the ALD oxide/graphene interface by having arrested the substrate-assisted ALD oxide growth in the nucleation stage. Importantly, via STEM we can also readily identify and quantify the presence of adventitious carbon contamination[42–44] at and near the non-2D/2D interfaces, which is otherwise hard to clarify. Adventitious carbon contamination is a ubiquitous factor in realistic and scalable (non-ultra-high-vacuum (non-UHV)) processing of 2D materials but its effect on oxide ALD on 2D materials still remains largely unexplored,[45–47] in particular at the atomic scale.

With our approach we confirm at high resolution participation of the underlying Cu catalyst in the substrate-assisted ALD process and confirm the role[12–15] of surface impurities like surface steps on ALD $HfO_2$ nucleation. Via emulation of annealing treatments via electron beam (e-beam) induced *in situ* crystallization,[48] we evidence the crystallization of the initially amorphous ALD $HfO_2$ to nanocrystalline $HfO_2$ of non-equilibrium cubic and/or tetragonal phase. Beyond this, our data suggests a key role of the 2D material's ever present adventitious carbon contamination on the subsequent ALD metal oxide nucleation. In particular our data points to adventitious carbon contamination acting as an unintentionally present seeding layer for ALD oxide nucleation i.e. we suggest an adventitious carbon contamination mediated nucleation mechanism of ALD oxides on graphene under our conditions. Given that adventitious carbon contamination is hard to avoid in any scalable materials processing this is a critical factor to consider in future ALD on 2D materials recipe design. Combined our work thereby resolves at the atomic scale several key influences underlying scalable ALD oxide growth and processing on 2D materials.

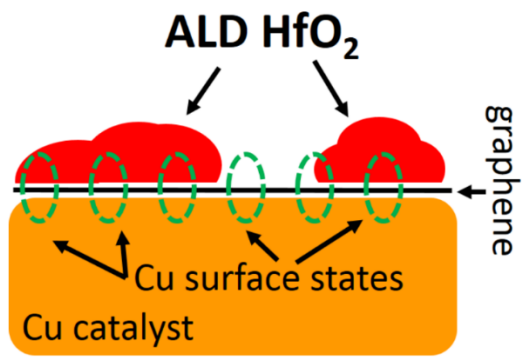

**Figure 1.** Schematic illustration of the hypothesized governing factors in substrate-assisted ALD[34] of HfO$_2$ on graphene.

Results and Discussion

We grow graphene by CVD on Cu catalyst foils[49,50] using conditions that lead to a continuous polycrystalline graphene film of predominantly monolayer graphene with an appreciable amount of bilayer and few layer graphene islands. This film structure allows us to readily study the dependence of our metal oxide deposition on the number of graphene layers. Substrate-assisted ALD[28,34] of HfO$_2$ is performed on the graphene/Cu foil stacks (i.e. the graphene remains in its as deposited state on its Cu catalyst during ALD) under typical ALD conditions using tetrakis(dimethylamido) hafnium (TDMAHf) and deionized water (H$_2$O) at a sample temperature of 200 °C.[28] We use up to 16 cycles of ALD to arrest the HfO$_2$ growth in the nucleation stage based on previous process calibrations (1 ALD cycle represents ~0.1 nm nominal oxide thickness).[28] We note that substrate-assisted ALD on the as grown graphene avoids any influence of residues[35,51,52] from polymer-assisted transfer processes.[12,27,36] After ALD the HfO$_2$/graphene stacks are released from the Cu catalyst foils and transferred as suspended membranes onto TEM grids using a polymer-free direct transfer method,[40] again avoiding detrimental effects of polymer residues on later imaging. The ALD HfO$_2$/graphene stacks are then studied using complimentary aberration corrected STEM (60 kV electron acceleration voltage)[41] and bright-field (BF) and dark-field (DF) transmission electron microscopy (TEM, 80 kV, incl. selected area electron diffraction (SAED)).[53] More details on experimental methods can be found in the methods section below.

Fig. 2 shows high angle annular dark field (HAADF) STEM data of ALD HfO$_2$ nucleations (16 ALD cycles) on CVD monolayer graphene at various magnifications. The intensity of HAADF STEM data scales linearly with specimen thickness (for a given material) and also provides materials contrast which is dependent[41] on atomic number Z with Z$^{\sim 1.64}$. Combined with the clear identification of monolayer graphene regions by resolving the graphene lattice[54] (inset in Fig. 2c) we identify the bright HAADF intensity regions in Fig. 2a as HfO$_2$ clusters ($Z_{Mo}$=72, $Z_O$=8) that nucleated on monolayer graphene (dark HAADF intensity regions, $Z_C$=6). Fig. 2b shows a corresponding false color coded recalculation of the HAADF intensity from Fig. 2a where the intensity has been normalized relative to a monolayer graphene layer (i.e. a graphene monolayer has relative intensity of 1). The observation of nucleation of the HfO$_2$ in clusters implies a Volmer-Weber-type growth mode of the ALD HfO$_2$. The graphene lattice is well resolved in Fig. 2c and does not show extended defects, suggesting excellent preservation of the graphene quality during the ALD process (as also corroborated by the SAED data for as deposited film in Fig. 3c below).

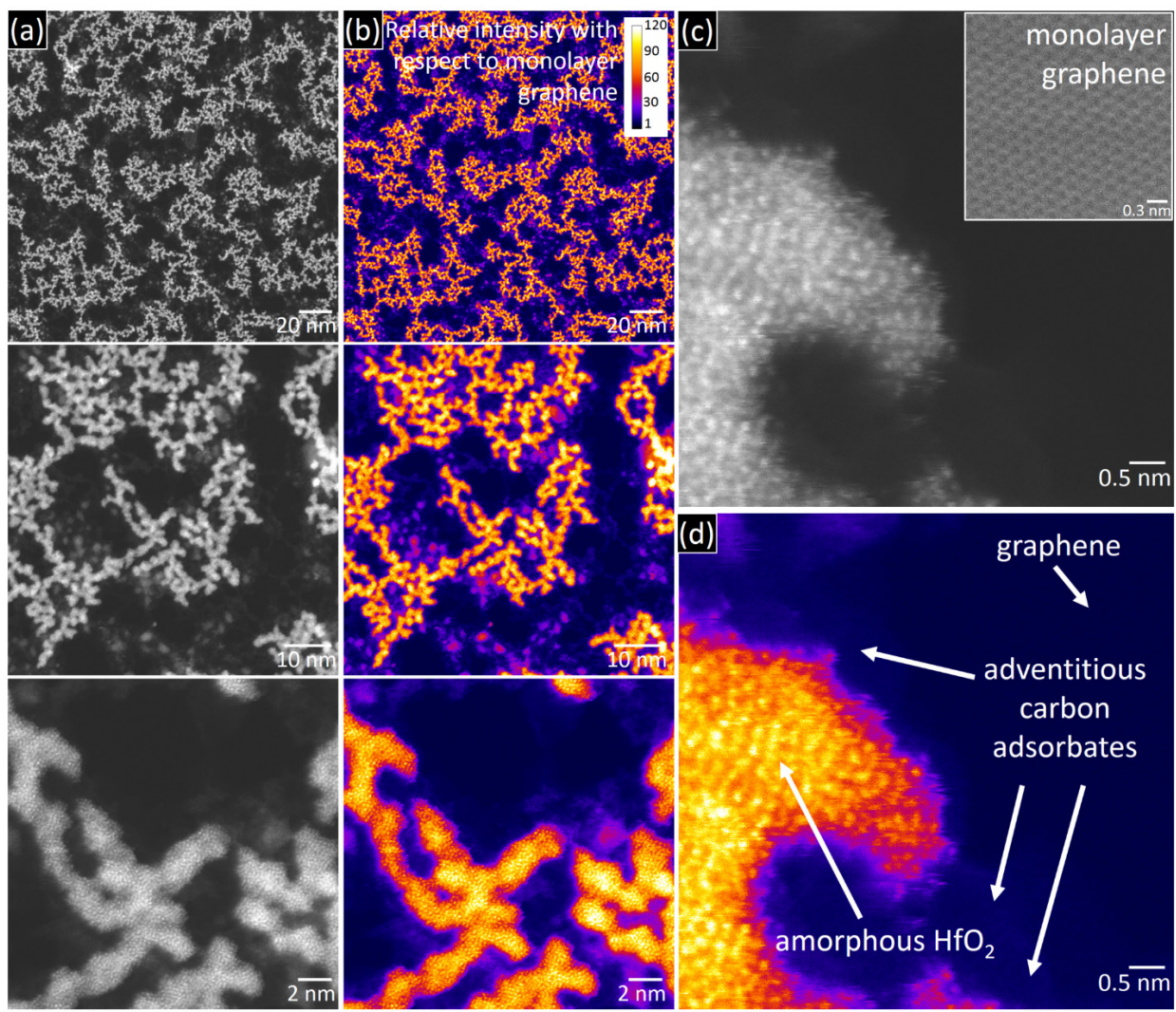

**Figure 2.** (a) HAADF STEM images of HfO$_2$ deposits from substrate-assisted ALD arrested in the nucleation stage (16 cycles) on monolayer CVD graphene at various magnifications (after release of the HfO$_2$/graphene stack from the Cu catalyst and transfer onto a TEM grid[40]). For corresponding Fourier transform (FT) data see Supporting Fig. S1. (b) False-color coded recalculation of the data in (a) with the intensities normalized relative to the intensity of monolayer graphene. (c) Zoom-in into (a). The inset shows HAADF data at higher magnification confirming a monolayer graphene lattice. (d) False-color coded recalculation of the data in (c) normalized as in (b) and with the salient features observed in image labelled.

The high resolution STEM data in Fig. 2 reveals that the as deposited $HfO_2$ clusters are amorphous in structure which is also corroborated by the corresponding Fourier transform (FT) data in Supporting Fig. S1. We find that the energy input from the scanning e-beam[48,55,56] in STEM can be used to *in situ* crystallize these as deposited amorphous $HfO_2$ clusters towards nanocrystalline $HfO_2$ as shown in Figure 3a. Cluster morphology, amorphicity of the as deposited $HfO_2$ and the e-beam induced crystallization are also corroborated at a larger field of view in the TEM measurements in Fig. 3b (BF-TEM) and Fig. 3c (SAED). Fig. 3b also shows that for extended e-beam exposures the $HfO_2$ clusters not only crystallize but also restructure (for >40 min e-beam exposure in the TEM the graphene conversely degrades under our conditions from beam-related damage[57]). The radially integrated SAED in Fig. 3d and its qualitative matching to the various $HfO_2$ polymorphs[58] (Supporting Fig. S3) suggest that under our conditions the $HfO_2$ on graphene does not crystallize into the thermodynamically most stable monoclinic $HfO_2$ phase.[33,59–61] Instead the SAED pattern in Fig. 3c is best matched (based on presence/absence of peaks) with either cubic or tetragonal $HfO_2$ (Supporting Fig. S3).[58] Both cubic and tetragonal $HfO_2$ are thermodynamically metastable phases but both are technologically sought after as they offer higher dielectric constants (*k*-values) than monoclinic $HfO_2$.[33,59–61] We note that even after long crystallization periods (45 min) we observe randomly oriented nanocrystalline $HfO_2$ (see $HfO_2$-associated rings in SAED in Fig. 3c) on the comparably much larger (several µm)[48,50] graphene grains (note also the single six-fold discrete spot pattern for graphene in the SAED in Fig. 3c, which suggests the presence of one single graphene grain over the entire field of view in Fig. 3b). This rules out strong epitaxial mechanisms between graphene and HfO$_2$ under our conditions, as no signs of dominant HfO$_2$/graphene in-plane orientation relations[31,32] emerge for our e-beam induced crystallization.

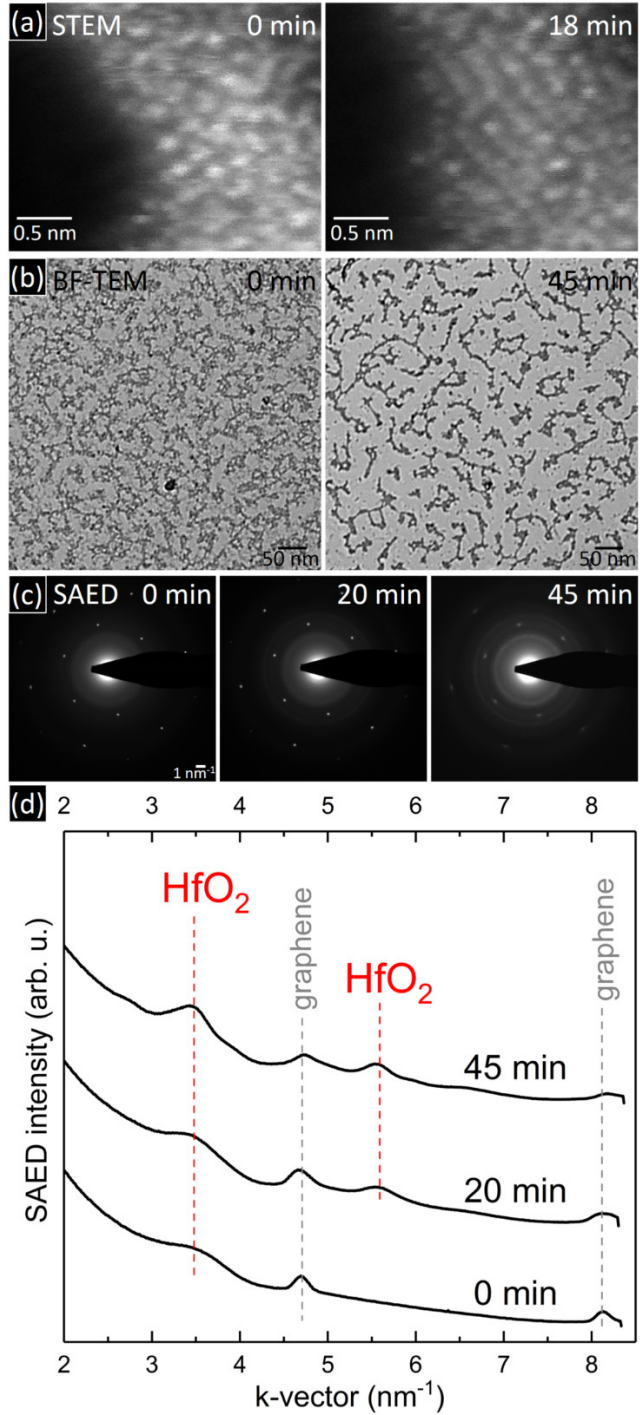

**Figure 3.** (a) HAADF STEM image of an ALD HfO$_2$ cluster (16 cycles) on graphene before extended e-beam exposure (0 min, left) and after 18 min of continuous e-beam scanning (right). (b) BF-TEM of HfO$_2$ clusters on graphene (16 cycles) before (left) and after 45 min e-beam

exposure (right). (c) SAED patterns corresponding to (b) after 0 min (left), 20 min (middle) and 45 min (right) e-beam exposure. The single discrete six-fold diffraction pattern is consistent with a graphene lattice and suggests the presence of only one single graphene grain in the entire field of view in (b). For confirmation of the monolayer nature of the graphene layer in (b) via SAED intensity analysis[62] see Supporting Fig. S2. The SAED data for the 0 min e-beam exposed film is consistent with monolayer graphene of high crystalline quality, confirming that the $HfO_2$ ALD coating leaves the graphene lattice intact. (d) SAED profiles radially integrated from (c) and with peaks of the identified phases labelled. For a qualitative phase match of the measured SAED data to the various polymorphs[58] of $HfO_2$ see Supporting Fig. S3.

Returning to investigating the $HfO_2$ in its as deposited state, an important observation from Fig. 2 is that the presence of $HfO_2$ is spatially correlated with the presence of an adventitious carbon contamination layer, as highlighted in Fig. 2d. As polymer residues are completely avoided in our processing scheme, the adventitious carbon contamination is related to sample exposure to ambient atmosphere during processing.[42–44] In our data no $HfO_2$ is found directly on or directly adjacent to atomically clean, bare graphene areas but $HfO_2$ is always surrounded by (and possibly placed on) adventitious carbon contamination regions. The linear thickness contrast combined with the element specificity in STEM[41] readily allows to identify the presence of such adventitious carbon that is ubiquitously present[42–44] in scalable (i.e. non-UHV) processing of materials. The observed spatial correlation of the $HfO_2$ nuclei with the adventitious carbon adsorbate implies two possible generic scenarios: (i) "$HfO_2$ attracts carbon": In this scenario, first ALD $HfO_2$ would nucleate on initially atomically clean graphene. Then the metal oxide preferentially would attract adventitious carbon from subsequent post-ALD ambient atmosphere sample handing and/or post-ALD from carbon residues in non-fully hydrolyzed TDMAHf precursor. (ii) "Carbon attracts $HfO_2$": In this scenario, adventitious carbon contamination from ambient atmosphere graphene handling is already present on the graphene before ALD. During ALD actually this pre-ALD adventitious carbon preferentially triggers ALD $HfO_2$ nucleation on the contamination. The latter scenario (ii) thereby implies an *adventitious carbon contamination mediated nucleation* of the ALD metal oxide on the 2D material.

We gain further insights to elucidate this potential role of the atmospheric adventitious carbon contamination in ALD oxide nucleation as well as other key factor governing ALD oxide nucleation on 2D materials by investigating the ALD $HfO_2$ nucleation over the transition from a graphene bilayer region to a graphene monolayer region in Figs. 4a,b.

We note that e-beam induced restructuring of adventitious carbon during high resolution STEM imaging can lead to complex phenomena of contamination dewetting and/or attraction of additional contamination under the beam[42,43] and can thus complicate quantitative conclusions about spatial correlations to the adventitious carbon locations. Throughout the STEM imaging sessions for the here presented samples, e-beam induced changes to the spatial distribution of the adventitious carbon were however minimal and the adventitious carbon contamination appeared as generally static, allowing us to here assess spatial correlations.

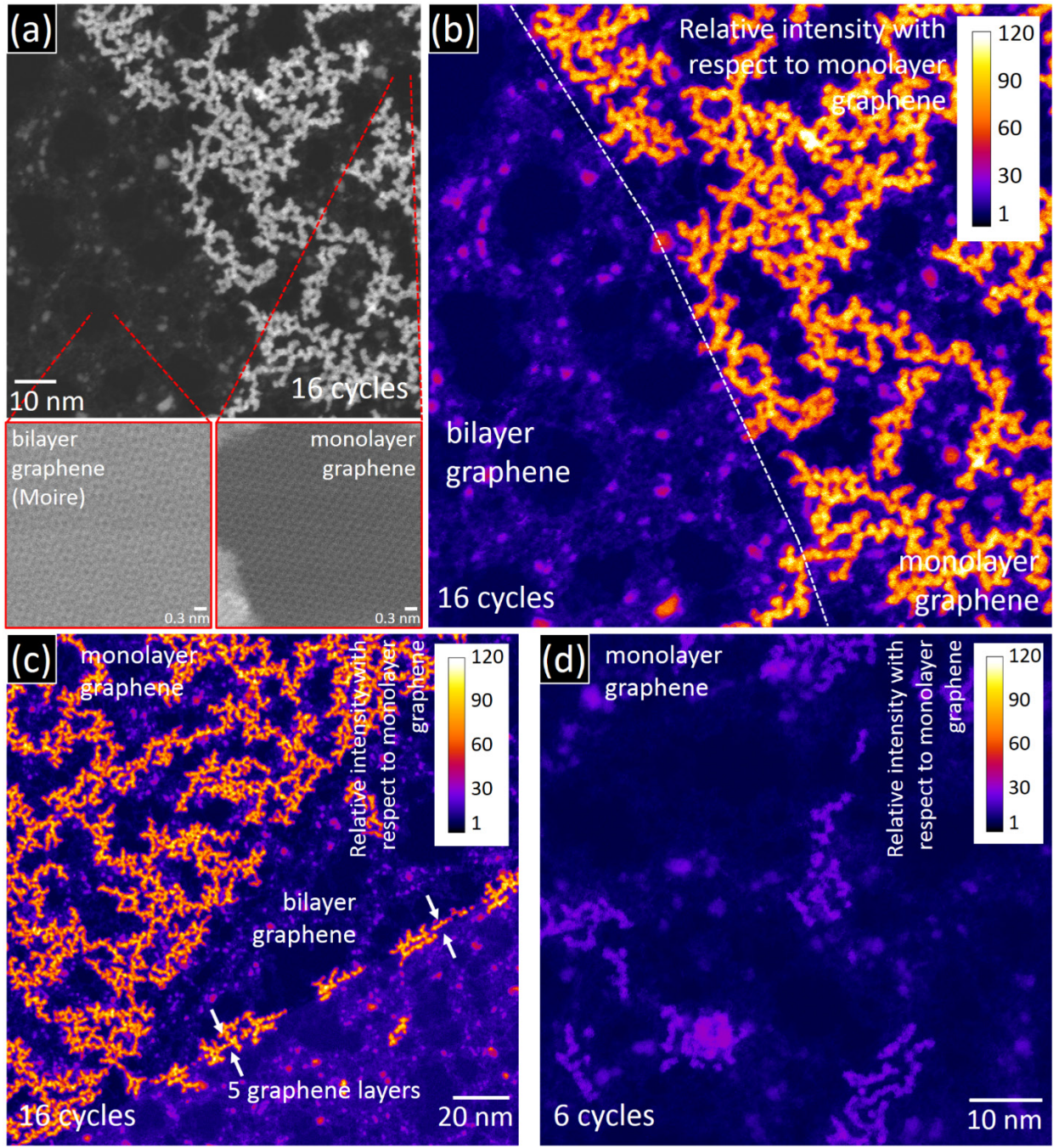

**Figure 4.** (a) HAADF STEM image of ALD HfO$_2$ deposits (16 cycles) over the transition from a graphene bilayer region (left) to a graphene monolayer region (right). The insets show HAADF data at higher magnification confirming the turbostratically stacked bilayer graphene lattice (left) and the monolayer graphene lattice (right), respectively. (b) False-color coded recalculation of the data in (a) with the intensities normalized relative to the intensity of monolayer graphene. (c) False-color coded HAADF STEM data (normalized as in (b)) of ALD HfO$_2$ deposits (16 cycles) over the transition from a monolayer graphene region (left) to a bilayer graphene region (middle) and further to a 5 few layer graphene region (right). (d) False-color coded HAADF STEM data (normalized as in (b)) of a reduced cycle number of ALD HfO$_2$ deposits (6 cycles) on a monolayer graphene region.

First, Fig. 4a,b shows that HfO$_2$ deposition is more pronounced on monolayer graphene. Quantifying the areal coverage of HfO$_2$ clusters via STEM intensity (threshold analysis) we measure a fractional areal HfO$_2$ coverage $\theta_{oxide, monolayer}$ of ~54% on the monolayer graphene region in Fig. 4b compared to only $\theta_{oxide, bilayer}$ ~20% HfO$_2$ areal coverage on the bilayer graphene regions. Similarly in Fig. 4c for steps from monolayer graphene to bilayer to few layer graphene we find comparatively high coverage on the monolayer graphene but much reduced coverage on the bi- and few layer regions. The scanning electron microscopy (SEM) data in Supporting Fig. S4 and the TEM data in Supporting Figure S5 further confirm these observations at larger fields of view. This dependence of HfO$_2$ growth on the number of graphene layers in our substrate-assisted ALD process affirms the initial assertion from above that the Cu catalyst under the graphene and its proximity to the oxide growth surface are of key importance in substrate-assisted ALD. Our observations are fully consistent with a model where the electronic properties of the Cu catalyst that emanate through monolayer graphene[39] partake in the ALD process due to the ultralow thickness of the graphene monolayer.[34] For graphene monolayers this contribution of the Cu surface states leads to improved coverage in the HfO$_2$ deposition on the otherwise hard-to-coat graphene surface.[28,34] When the distance between Cu catalyst and HfO$_2$ growth front is however doubled for bilayer graphene (and further increased for higher graphene layer numbers) the contribution of the underlying Cu surface states to the ALD reactions are strongly reduced[39] thereby negating the gain from the Cu-substrate assistance and thus resulting in much lower $HfO_2$ coverage on bi- and few layer graphene, similar to prior reports of depositions on graphite (i.e. no Cu underneath at all).[28,34]

As a second point, Fig. 4c also highlights the role of surface irregularities[8,9,12–15,17] like steps or folds between graphene layers on ALD nucleation, as we find strongly increased $HfO_2$ deposition along the step from the bi- to the few-layer graphene region (marked by white arrow). Such preferential ALD oxide nucleation along steps has been previously ascribed[8,9,12–15] to the (compared to the inert graphene basal plane) much more chemically reactive sites at graphene layer edges. The higher reactivity at the layer edges thereby leads to preferential adsorption of ALD precursors and/or preferential nucleation of ALD oxides. Prior work has also identified grain boundaries in polycrystalline monolayer graphene films to act as preferential nucleation sites for ALD due to their higher reactivity.[12,13,63] As we show in Supporting Fig. S6 via DF-TEM over a grain boundary in our CVD monolayer graphene film, we find however no indications of preferential ALD $HfO_2$ nucleation along the grain boundary in comparison to the graphene basal plane for our conditions. We note however that on both grain boundaries[52] and basal plane[42–44] are typically covered by adventitious carbon and for our conditions ALD nucleation may be primarily adventitious carbon contamination mediated, which we further examine now in detail.

The third observation in Fig. 4a,b is that the fractional area of atomically clean graphene ($\theta_{graphene}$, showing lattice contrast without any signs of adventitious carbon adlayers) remains roughly constant across the step as we move from the monolayer graphene region (high $HfO_2$ coverage) to the bilayer graphene region (low $HfO_2$ coverage), while the $HfO_2$ coverage drastically changes. Fractional area of atomically clean graphene areas is a measure of cleanness of the graphene. Quantification affirms the visual impressions that the graphene is approximately equally clean on the monolayer and bilayer regions, as we find an areal fraction $\theta_{graphene,\ monolayer}$ ~17 % of bare graphene regions on the monolayer compared with only slightly increased bare graphene areas ($\theta_{graphene,\ bilayer}$ ~22%) on the bilayer region. We note that the sum of the fractional areas of regions with adventitious carbon contamination *and* $HfO_2$ oxide coating remains roughly constant across the step ($\theta_{contamination} + \theta_{oxide} = 1 - \theta_{graphene}$; although note that ALD $HfO_2$ deposition can only have occurred on the "top" surface of the graphene/Cu stacks due to the substrate-assisted ALD while post-ALD adventitious carbon may deposit on both sides of the suspended $HfO_2$/graphene membranes on the TEM grids). Due the visual spatial continuity of the adventitious carbon network throughout Figs. 2 and 4 a plausible assumption to make is that the adventitious carbon adsorbates are a continuous layer underneath the $HfO_2$ deposits. This assumption leads to the interpretation of our approximately constant $\theta_{graphene}$ values irrespective of $HfO_2$ coverage as the graphene being similarly covered by adventitious carbon on the mono- and bilayer regions irrespective of amount of ALD $HfO_2$ deposited. This interpretation, in connection with the absence of $HfO_2$ directly on the bare graphene, then implies that the $HfO_2$ growth follows our scenario (ii) "Carbon attracts $HfO_2$" from above i.e. the ALD $HfO_2$ gets only preferentially deposited on pre-existing adventitious carbon deposits that are already present pre-ALD. (If scenario (i) "$HfO_2$ attracts carbon" was dominating the amount of adventitious carbon deposits should directly scale with the amount of $HfO_2$ deposits, contrary to what we observe.)

Further evidence supporting the "Carbon attracts $HfO_2$" scenario is given in Fig. 4d where we have reduced the number of ALD from 16 (as in Fig. 1 and Fig. 4a-c) to 6 cycles and thus have reduced the amount of HfO$_2$ deposited. Consistently, we find a lower amount of HfO$_2$ in the STEM images ($\theta_{oxide,\ 6\ cylces}$ ~15% areal coverage; maximum HAADF intensity from 6 cycles HfO$_2$ clusters reduced by a factor ~4 implying ~4× thinner HfO$_2$ clusters compared to the 16 cycles in Fig. 2 and Fig. 4a,b on the graphene monolayers). Importantly, while the amount of HfO$_2$ deposited thereby scales approximately with number of ALD cycles, the fractional bare graphene area does not but instead remains with $\theta_{graphene,\ 6\ cycles}$ ~24% approximately the same as for the 16 cycles ALD in Fig. 1 and Fig. 4a-c. Making again the assumption that the HfO$_2$ deposits are placed on adventitious carbon we thereby again find a similar contamination level of the graphene as in the 16 cycles depositions (Fig. 2 and Fig. 4a,b on the graphene monolayers) despite the much reduced HfO$_2$ deposition. This corroborates that HfO$_2$ preferentially nucleated on the pre-existing adventitious carbon contamination. We note that also non-ALD-treated graphene that underwent similar storage and transfer conditions shows bare graphene area fractions of $\theta_{graphene,\ no\ ALD}$ ~25-30% with the remaining area covered by adventitious carbon deposits. Such coverage values of non-UHV processed graphene are also fully consistent with previous literature on adventitious carbon deposition, which showed that adsorption of adventitious carbon readily and very swiftly occurs when samples are exposed to ambient.[42,43,64] Combined, our observations strongly suggest that adventitious carbon is present before ALD of HfO$_2$ and that this contamination mediates the locations of HfO$_2$ nucleation i.e. we observe an adventitious carbon contamination mediated nucleation of the ALD HfO$_2$ on the graphene films.

A suggested mechanism behind this role of the adventitious carbon contamination stems from their chemical nature: Unlike the inert graphene basal plane which is void of sites for chemical binding for the ALD precursors, adventitious carbon consists of a mixture of sp$^2$- and sp$^3$-hybridised (hydro-)carbon deposits which include a large a variety of available binding sites and surface groups such as –OH, –COOH etc.[65] Such surface sites are much more reactive towards adsorption of ALD precursor and oxidant molecules than the graphene basal plane,[66] thus facilitating ALD oxide nucleation. Therefore compounds that have similar surface groups as adventitious carbon (e.g. derivates of perylene[16,19], benzyl alcohol,[17] pyrene,[17] graphene oxide[20,22] etc.) have been deliberately used in previous work as chemical seeding layers to improve and control ALD nucleations on graphene and other graphitic materials. We now suggest that the intrinsically present atmospheric adventitious carbon contamination on scaleably processed graphene (as employed here) presents itself as an unintentionally present seeding layer for ALD oxide nucleation. This is even true when influences of additional carbonaceous polymer residues from graphene transfers are completely avoided (as here).

The here spatially resolved observed influence of adventitious carbon adsorbates on $HfO_2$ ALD is consistent with prior integral chemical fingerprinting work by X-ray photoelectron spectroscopy of the influence of adventitious carbon contaminations on oxide ALD on graphite[45,46] and $MoS_2$.[47] Also prior TEM work on metal oxide growth by ALD[12,14,30] as well as metal[43] and metal-oxide[67] deposition by evaporation found deposits preferentially on carbon adsorbates (be they processing-related polymer residues or adventitious contamination), in line with our observations. A corollary from our hypothesis of adventitious carbon acting as an unintentional ALD seeding layer is that controlled exposure to ambient atmosphere conditions pre or during ALD should impact on the ALD oxide nucleation results. In particular, repeated exposure to ambient during the ALD coating process can be surmised to lead to incremental consecutive coating of remaining clean graphene areas by adventitious carbon which in turn for the next cycles of ALD would be expected to act as ALD oxide seeding layers. This would imply that repeated interruption of the ALD process by ambient air exposures should lead to more homogeneous ALD oxide coating on graphene. Interestingly, recent literature[68] has reported exactly this suggested effect of improving ALD oxide coating homogeneity by intermitted air exposure during ALD cycling for ALD $Al_2O_3$ on graphene (albeit without identification of the underlying mechanism). This previous work[68] thereby aligns very well with the here identified key role of adventitious carbon contamination on ALD oxide nucleation on 2D materials.

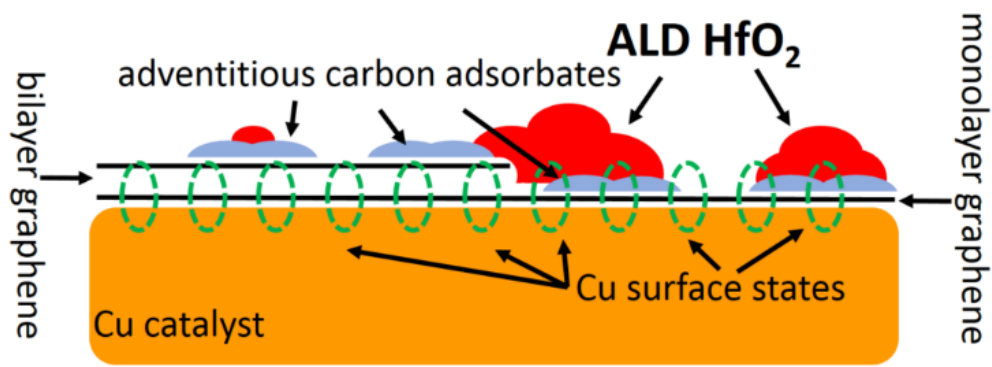

**Figure 5.** Revised schematic illustration of the governing factors in substrate-assisted ALD[34] of $HfO_2$ on graphene, based on the findings in this study. Besides the here confirmed graphene-layer-dependent participation of the Cu support in the ALD process and the effect of surface irregularities such as steps on ALD oxide nucleation, particularly the here suggested adventitious carbon contamination mediated ALD oxide nucleation mechanism is highlighted in the sketch.

**Conclusions**

In summary, we have employed high resolution electron microscopy techniques to investigate the nucleation stage in ALD of the important metal-oxide $HfO_2$ on the important 2D material graphene. Fig. 5 summarizes our key findings: Particular to the here employed substrate-assisted ALD process we confirm the active and graphene-layer-dependent role of the underlying Cu catalyst during metal oxide ALD on the Cu supported graphene. Beyond this particular ALD process our data confirms the role of surface irregularities such as steps on nucleation. We also find via e-beam induced crystallization experiments that the thin, initially amorphous ALD $HfO_2$ crystallizes to non-equilibrium cubic/tetragonal $HfO_2$ under our conditions. Finally our data suggests a key role of atmospheric adventitious carbon contamination on the graphene on the subsequent ALD metal-oxide nucleation even if extraneous polymer residues from graphene transfers are fully avoided. We suggest that the increased chemical reactivity of adventitious carbon adsorbates compared to the inert basal plane of graphene leads to these adsorbates acting as an unintentional seeding layer for metal/oxide ALD nucleation on graphene under conditions. As adventitious carbon contamination is ubiquitous in scalable 2D materials processing, this is a critical finding to take into account in future recipe development for 2D materials coating by ALD.

**Methods**

Continuous polycrystalline monolayer graphene films with some bilayer and few layer graphene islands were grown by CVD in a 100 mm hot-walled tube furnace (Graphene Square) on Cu catalysts (Alfa Aesar, 25μm thick, 99.999% purity).[49,50,64] Prior to growth, these catalysts were annealed for 90 min in $H_2$/Ar mixture (1:4 ratio) at a temperature of ~1050°C. Graphene growth was subsequently carried out at the same temperature using $H_2$ diluted $CH_4$/Ar mixture (1:1000 ratio) at a $CH_4$ partial pressure of $10^{-3}$ mbar for 60 min. Storage time in ambient air between CVD growth and subsequent ALD was ~1 day. The substrate-assisted ALD of $HfO_2$ directly on the CVD graphene/Cu foil stacks was then carried out in a Cambridge Nanotech Savannah S100 G1 system at a substrate temperature of 200°C using tetrakis(dimethylamido) hafnium (TDMAHf, purity >99% Sigma Aldrich) volatized at 80°C as the metal precursor and deionized water ($H_2O$) volatized at 40°C as the oxidant in a pretreatment mode.[28,34,38] Prior to the deposition, the ALD chamber was evacuated to a base pressure of $~5\times10^{-1}$ Torr and purged with 10 min of $N_2$ at a flow rate of 20 sccm. The pretreatment was carried out by 10 pulses of $H_2O$, each at ~0.15 Torr·s and was separated by 12 s of $N_2$ purge. This was then followed by ALD cycles without breaking the vacuum for a total of 2–16 cycles. In each ALD cycle, ~0.15 Torr·s TDMAHf and ~0.2 Torr·s $H_2O$ were delivered alternatively and separated by 20 s of $N_2$ purge. The metal oxide/graphene stacks were then released from the Cu catalyst using $FeCl_3$-based aqueous wet etches and suspended as membranes using a polymer-free transfer process[40] onto holey carbon TEM grids with regular hole arrays (Quantifoil). Note that in the transfer process the $FeCl_3$ wet etch only came into contact with the $HfO_2$/graphene stacks on their graphene underside i.e. the $FeCl_3$ etch did not come in contact with the $HfO_2$ deposits on top of the stack.

Both the substrate-assisted ALD process and the employed direct transfer method after ALD avoid persistent polymer residues[35,51,52] typically associated with polymer-based transfers. Throughout and after fabrication samples were stored and transported in ambient air.

STEM was measured in an aberration corrected Nion UltraSTEM 100 at an electron acceleration voltage of 60 kV.[41] We simultaneously acquired HAADF (80 to 200 mrad) and MAADF (40 to 80 mrad) signals. Vacuum at the sample in the STEM was $\sim 10^{-9}$ mbar to minimize reactions with residual gas species. Beam currents in the STEM are $\sim 30$ pA for $\sim 1$ Å$^2$ spot sizes, resulting in electron dose rates under the beam of $\sim 5 \times 10^8$ e$^-$Å$^{-2}$s$^{-1}$. This in turn translates to average dose rates of $\sim 5 \times 10^4$ e$^-$Å$^{-2}$s$^{-1}$ for continuous scanning of 10 nm × 10 nm areas. For *in situ* crystallization in STEM (Fig. 3a) continuous e-beam exposure was achieved via continuous STEM scanning at fields of view of 5 × 5 nm$^2$ ($\sim 2 \times 10^5$ e$^-$Å$^{-2}$s$^{-1}$). We note that scanning over larger fields of view (and correspondingly lower average dose rates) did not lead to observable HfO$_2$ crystallization but imaged the samples modification-free. Samples were annealed prior loading into the STEM at $\sim 140$ °C in $10^{-5}$ mbar overnight to desorb lightly bound adventitious hydrocarbons and adsorbed water from sample storage in ambient. We crosschecked with TEM and SAED (without any pre-loading anneal) that such mild annealing before STEM loading did not lead to HfO$_2$ modifications. TEM (BF and DF)[53] and SAED was measured in a Philips CM200 TEM employing 80 kV electron acceleration voltage. In TEM the sample rests in a vacuum of $\sim 10^{-6}$ mbar. A wide e-beam was used for TEM imaging and SAED at electron dose rates ($\sim 4 \times 10^1$ e$^-$Å$^{-2}$s$^{-1}$) that did not induce HfO$_2$ crystallization. To observe HfO$_2$ crystallization in the TEM the e-beam had to be focused to achieve electron dose rates of $\sim 3 \times 10^3$ e$^-$Å$^{-2}$s$^{-1}$. SEM was measured in a Zeiss Sigma 55VP SEM at 2 kV electron acceleration voltage using an in-lens detector.

SUPPORTING INFORMATION

Supporting Information (additional electron microscopy data and analysis) is available below.

ACKNOWLEDGMENTS

B.C.B acknowledges funding from the European Union's Horizon 2020 research and innovation program under the Marie Skłodowska-Curie Grant Agreement 656214-2DInterFOX. J.C.M. acknowledges the Austrian Science Fund (FWF) under project P25721-N20. S.H. acknowledges funding from EPSRC (EP/K016636/1). A.I.A acknowledges funding from the European Union's Horizon 2020 research and innovation program under the Marie Skłodowska-Curie Grant Agreement 645725-FRIENDS2.

Supporting Information to:

# Resolving the Nucleation Stage in Atomic Layer Deposition of Hafnium Oxide on Graphene

Bernhard C. Bayer,[1,2,*] Adrianus I. Aria,[3] Dominik Eder,[1] Stephan Hofmann,[4] Jannik C. Meyer[2,#]

[1]Institute of Materials Chemistry, Vienna University of Technology (TU Wien), Getreidemarkt 9, A-1060 Vienna, Austria

[2]Faculty of Physics, University of Vienna, Boltzmanngasse 5, A-1090 Vienna, Austria

[3]Surface Engineering and Precision Institute, School of Aerospace, Transport and Manufacturing, Cranfield University, College Road, Cranfield, MK43 0AL, UK

[4]Department of Engineering, University of Cambridge, 9 JJ Thomson Avenue, Cambridge, CB3 0FA, UK

[*]Corresponding Author e-mail: bernhard.bayer-skoff@tuwien.ac.at

[#]Present address: Institute for Applied Physics, Eberhard Karls University of Tuebingen, Auf der Morgenstelle 10, D-72076 Tuebingen, Germany and Natural and Medical Sciences Institute at the University of Tuebingen, Markwiesenstr. 55, D-72770, Reutlingen, Germany

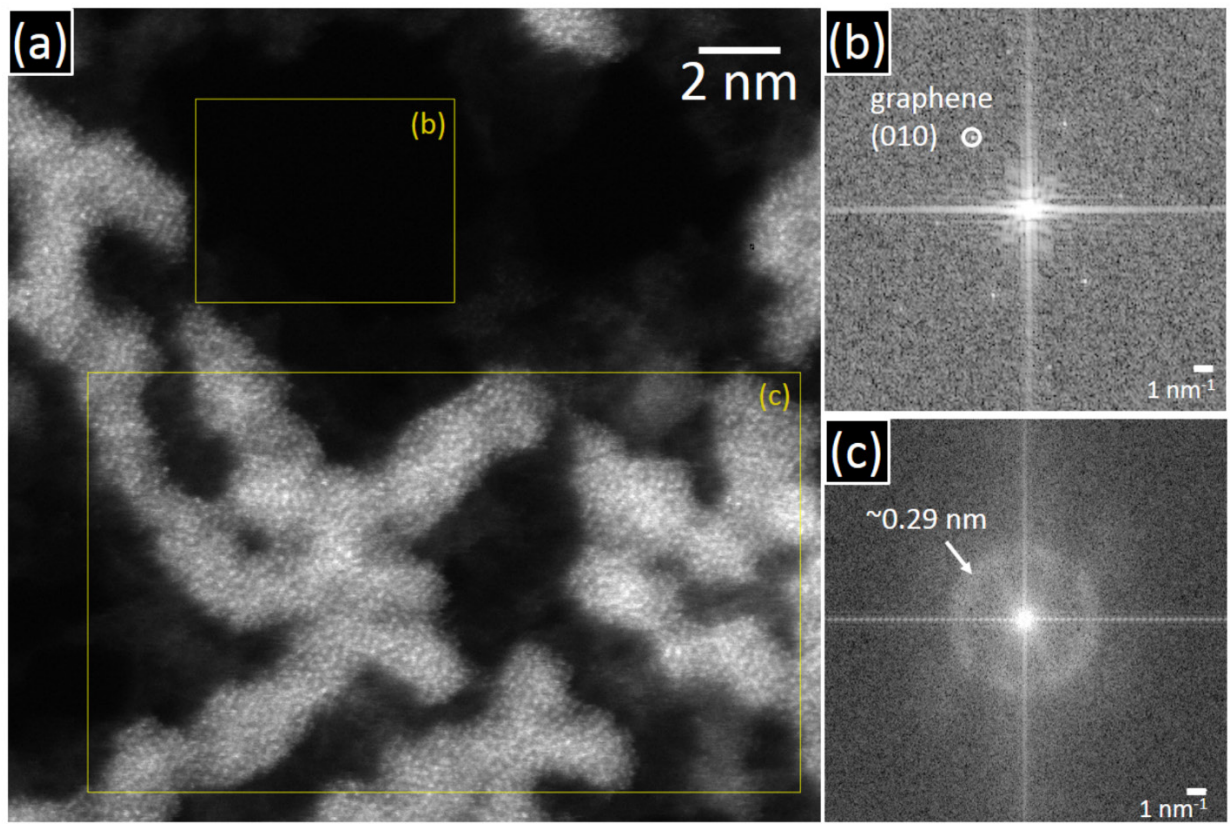

**Supporting Figure S1.** (a) Medium angle annular dark field (MAADF) scanning transmission electron microscopy (STEM) image of $HfO_2$ deposits from substrate-assisted atomic layer deposition (ALD) (16 cycles) on chemical vapor deposited (CVD) graphene (corresponding to high angle annular dark field (HAADF) data in Fig. 2a/bottom panel from main text). (b) Fourier transform (FT) of the correspondingly marked region in (a) of the atomically clean, bare graphene area. The FT clearly shows the six-fold symmetry and lattice distances consistent with a graphene lattice.[S1] (c) FT of the correspondingly marked region in (a) of the $HfO_2$ clusters. The FT shows a broad halo centered ~0.29 nm. This is consistent with an amorphous $HfO_2$ deposit.[S2]

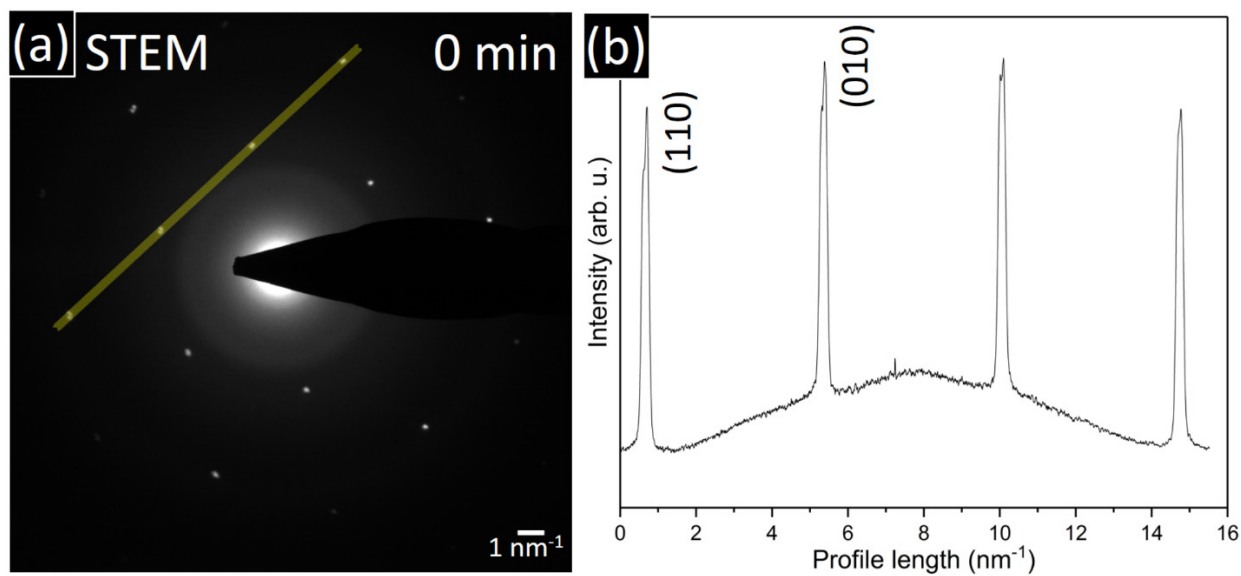

**Supporting Figure S2.** (a) Selected area electron diffraction (SAED) of $HfO_2$ clusters on graphene (16 cycles substrate-assisted ALD) before electron beam (e-beam) illumination (0 min, replot of Fig. 3c/left panel from main text). (b) SAED intensity profile extracted along the yellow line marked in (a), which via SAED intensity analysis[1] confirms the monolayer nature of the graphene layer investigated in Fig. 3b-d.

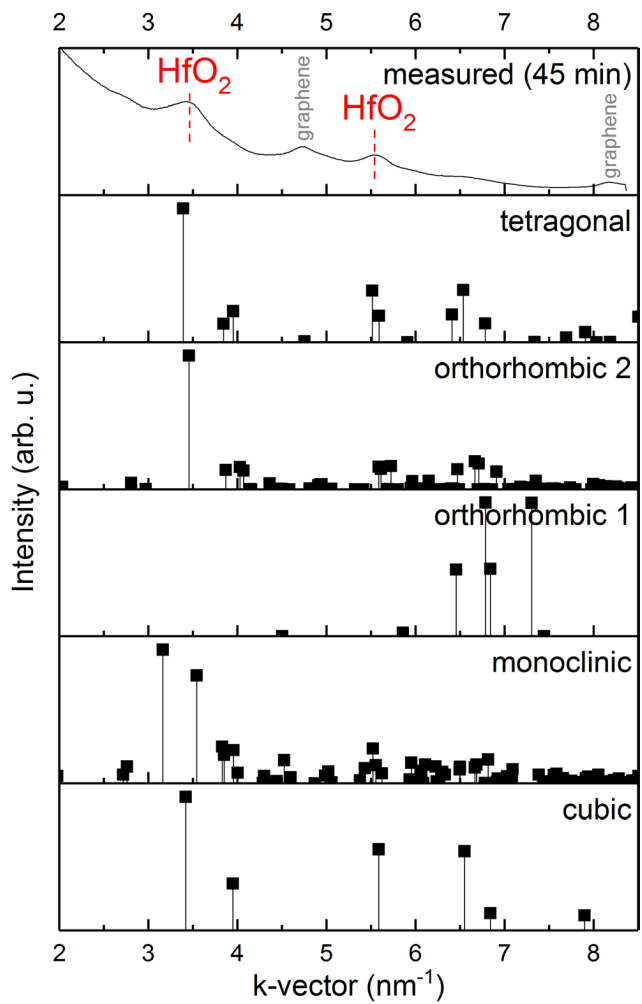

**Supporting Figure S3.** The top panel shows a measured radially integrated SAED profile for ALD HfO$_2$ deposits on graphene after 45 min e-beam illumination and *in situ* crystallization of the HfO$_2$ (replot from Fig. 3d/top profile) with peaks of the identified phases labelled. The panels below show kinematically calculated (Philips X'Pert Plus) powder diffraction patterns of the various polymorphs of HfO$_2$ for a qualitative phase identification of the HfO$_2$ deposits. The following polymorphs of HfO$_2$ were considered (Inorganic Crystal Structure Database (ICSD) entry number):[S2] cubic (173967), monoclinic (173964), orthorhombic 1 (173968), orthorhombic 2 (173965), tetragonal (173966). For graphene the entry for ICSD graphite (53781) was used. Based on the presence/absence of reflections the best qualitative match to the measured SAED profile in the top panel is cubic and/or tetragonal $HfO_2$. Monoclinic $HfO_2$ is the thermodynamically most stable $HfO_2$ polymorph whereas cubic and tetragonal $HfO_2$ are only metastable non-equilibrium polymorphs.[S3–S6]

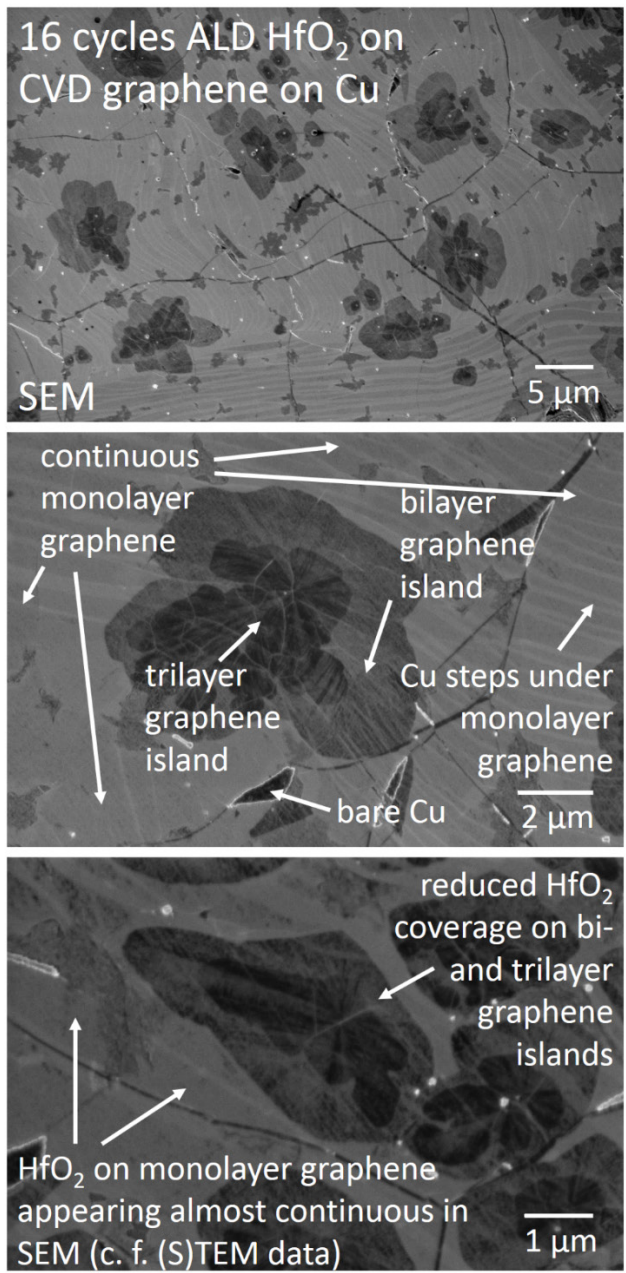

**Supporting Figure S4.** Scanning electron microscopy (SEM) images of HfO$_2$ deposits from substrate-assisted ALD arrested in the nucleation stage (16 cycles) on CVD graphene at various magnifications still remaining on the Cu catalyst (i.e. before release of the HfO$_2$/graphene stack from the Cu foil and transfer to transmission electron microscopy (TEM) grids). The CVD graphene is a continuous monolayer graphene film with a fraction of bilayer and few layer

graphene islands.[S7,S8] Salient features in the images are labelled according to the main text. ALD HfO$_2$ deposition is most pronounced on the monolayer graphene regions while HfO$_2$ deposits are reduced on bi- and trilayer graphene islands. We note that this 16 cycle ALD HfO$_2$ deposition was shown to consist of HfO$_2$ clusters by high resolution scanning transmission electron microscopy (STEM) in Fig. 2 in the main text, while in the here employed lower resolution SEM at a wider field of view (several μm) the HfO$_2$ already appears as a largely closed film with only small variations in HfO$_2$-associated contrast at the ridges of the Cu foil,[S9] suggesting that care has to be taken when evaluating oxide film homogeneity by SEM.

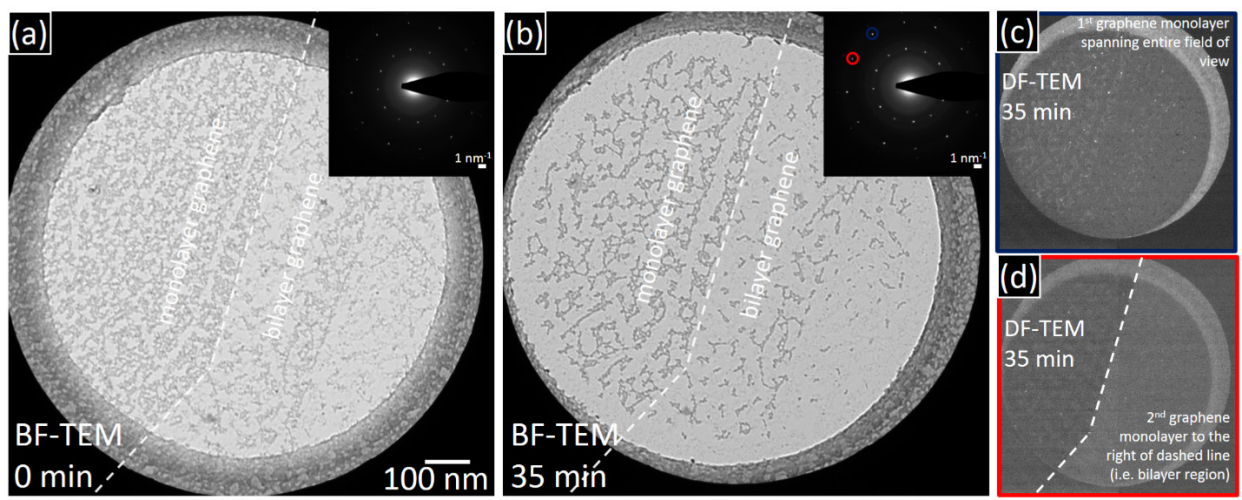

**Supporting Figure S5.** (a) Bright field (BF) TEM image of ALD $HfO_2$ deposits (16 cycles) over the transition from a graphene monolayer region (left) to a graphene bilayer region (right) before extended e-beam illumination (0 min). The assignment of the graphene layer numbers in the field of view stems from the dark field (DF) TEM analysis presented below. The inset shows the corresponding SAED pattern. (b) BF-TEM image of the same region as (a) after 35 min e-beam exposure in the TEM. The inset shows the corresponding SAED pattern. (c) DF-TEM image[S10] of the region corresponding to (b) imaged via the blue circled reflection in the SAED inset in (b), showing one graphene layer to span across the entire field of view. (d) DF-TEM image of the region corresponding to (b) imaged via the red circled reflection in the SAED inset in (b), showing that another graphene island is turbostratically stacked on top of the first layer in the right part of the field of view (i.e. a bilayer graphene region is visible in the right part of the field of view). Notably the $HfO_2$ coverage (dark intensity regions in BF-TEM in (a) and (b)) is consistently higher on the monolayer graphene region compared to the bilayer graphene region.

S8

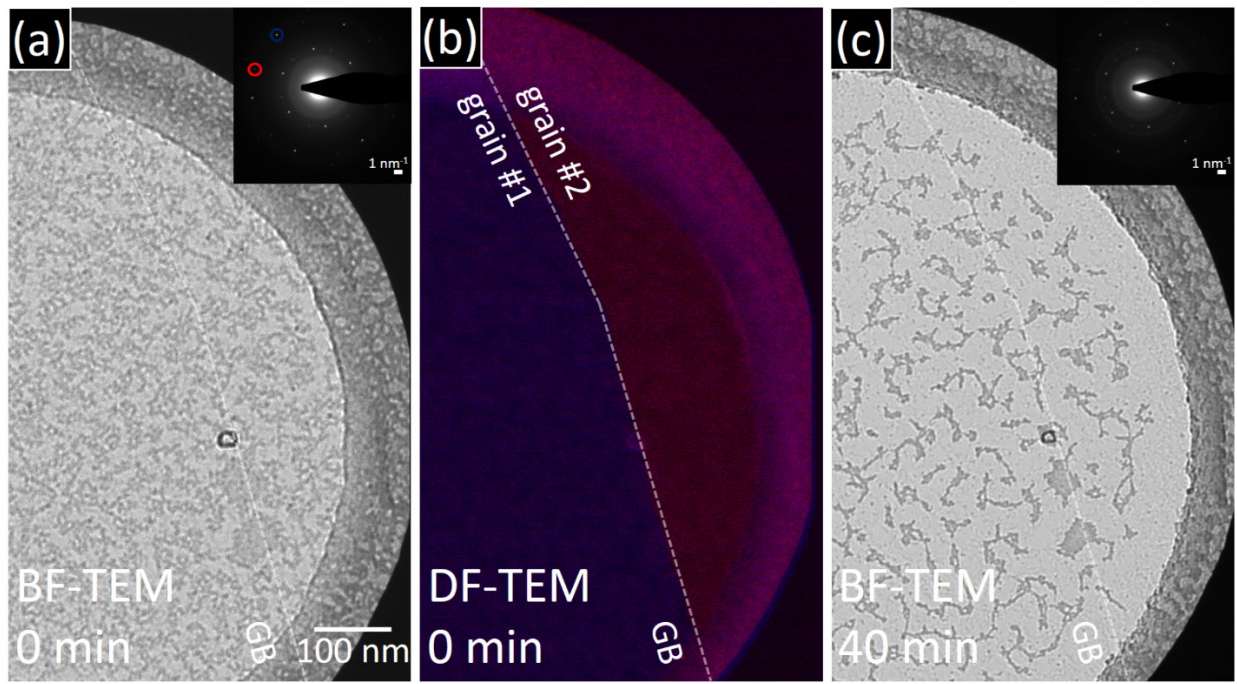

**Supporting Figure S6.** (a) BF-TEM image of ALD HfO$_2$ deposits (16 cycles) over a grain boundary (GB, marked by dashed white line) in the polycrystalline monolayer graphene film before extended e-beam illumination (0 min). The inset shows the corresponding SAED pattern, displaying two sets of six-fold graphene reflections with an rotational offset amongst the two sets, consistent with a grain boundary in the graphene monolayer film.[S10,S11] The identification of the location of the GB in the graphene stems from the DF-TEM analysis presented below. (b) False colored overlay of DF-TEM images of the same region as (a) acquired via the red and blue circled reflections in the SAED in the inset in (a) which reveals the location of two respective grains and their GB.[S10,S11] (c) BE-TEM image of the same region as (a) after 35 min e-beam exposure in the TEM. The inset shows the corresponding SAED pattern. Consistent with the other data, extended e-beam illumination led to HfO$_2$ restructuring and crystallization. Notably the HfO$_2$ coverage (dark intensity regions in BF-TEM in (a) and (c)) does not follow the location of the identified GB in any particular way (neither before nor after extended e-beam exposure). This indicates that under our conditions ALD HfO$_2$ nucleation is not preferential along graphene GBs compared to the graphene basal plane (albeit adventitious carbon contamination of both GB and basal plane has to be considered).